\documentclass[12pt]{article}
\usepackage{mathrsfs}
\usepackage{amssymb}
\usepackage{amsmath}
\usepackage[amsmath,thmmarks]{ntheorem}
\usepackage{cite}

\usepackage{graphicx}
\usepackage{dcolumn}
\usepackage{bm}
\usepackage{slashed}

\newcommand{\be}{\begin{eqnarray}}
\newcommand{\ee}{\end{eqnarray}}

\begin{document}

\begin{titlepage}

\makebox[6.5in][r]{\hfill ANL-HEP-PR-11-4}

 \makebox[6.5in][r]{\hfill nuhep-th/11-03}

\vskip.80cm
\begin{center}
{\Large {\bf Constraints on heavy colored scalars from Tevatron's Higgs exclusion limit}} \vskip.5cm
\end{center}
\vskip0.2cm

\begin{center}
{\bf Radja Boughezal}
\end{center}
\vskip 6pt
\begin{center}
{\it High Energy Physics Division, Argonne National Laboratory, Argonne, IL
  60439, USA} 
\end{center}

\vglue 0.3truecm

\begin{abstract}
\vskip 3pt \noindent
The null search for the Higgs boson at the Tevatron implies strong
constraints on heavy colored particles that increase the gluon-fusion
induced production rate of the Higgs.  We investigate the implications
of the Tevatron exclusion limit on example extensions of the
Standard Model that contain a new scalar state transforming as either
an adjoint or a fundamental under the QCD gauge group.  The bounds on the
adjoint (fundamental) scalar mass exceed 200 GeV (100 GeV) for
natural choices of scalar-sector parameters.

\end{abstract}

\end{titlepage}
\newpage

\section{Introduction \label{sec:intro}}

The hunt for the Higgs boson in order to uncover its role in
electroweak symmetry breaking has been actively undertaken during the
previous few years at the Tevatron accelerator complex.  
The CDF and D0 collaborations at the Tevatron have recently announced a 95\%
CL exclusion limit on a Standard Model (SM) Higgs boson with a mass in 
the ranges $100 \, \text{GeV} \leq m_h \leq 109 \, \text{GeV}$ and $158 \,
\text{GeV} \leq m_h \leq 175 \, \text{GeV}$~\cite{:2010ar}.  This
result is the culmination of intense efforts by experimentalists to
reduce systematic errors inhibiting the search and to devise
sophisticated multi-variate analysis techniques to enable control of
the severe backgrounds~\cite{Herndon:2008uv}, and by theorists to calculate precisely the
production rate and distribution shapes of the Higgs within the
SM~\cite{Boughezal:2009fw}.  Although the Tevatron will soon cease
operation and the search for the Higgs boson will shift to the LHC, there is much to still be learned from
Tevatron's null search.  The lack of an observed Higgs signal, and our
ability to calculate very precisely what is expected in the Standard
Model, implies significant constrains on additional 
new states that increase the Higgs cross section.  This has been demonstrated by CDF and D0 by their placement of stringent bounds on the existence of a fourth 
generation of fermions through its contribution to the gluon-fusion
production of a Higgs~\cite{Aaltonen:2010sv}.  In this work the
collaborations placed model-independent bounds on $\sigma_{gg \to h}
\times BR(h \to WW)$, allowing the implications of their null search
on other SM extensions which affect the partonic process $gg \to h \to
W^+W^-$ to be investigated.

Our goal in this manuscript is to further examine the implications of the Tevatron exclusion of the Higgs boson for physics beyond the Standard Model.  We consider 
two example states that contribute significantly to the gluon-fusion
production of a Higgs boson: color-fundamental and color-adjoint
scalars.  We calculate the gluon-fusion production cross section and
$W^+W^-$ decay width of the Higgs through next-to-next-to-leading-order (NNLO)
in QCD using an effective-theory approach, and use the results of
Ref.~\cite{Aaltonen:2010sv} to constraint the scalar parameter space.
We have previously presented a detailed discussion of how
color-adjoint scalars affect the Higgs production cross section~\cite{Boughezal:2010ry}, and we refer
to this work for many technical details.  For natural choices of
scalar parameters, we find bounds of one hundred to several hundred
GeV on the scalar mass, depending on both representation and the Higgs
mass.  While our interest  in these specific particles is because of their utility as phenomenological examples, we note that such states appear in extensions of the Standard Models.  
For example, adjoint scalars in the $(\bf{8},\bf{1})_{0}$ representation arise in theories with universal extra dimensions~\cite{Dobrescu:2007xf,Dobrescu:2007yp}. We 
note that the examination of how the Tevatron exclusion limit impacts modes of new physics has received other attention in the literature~\cite{Bechtle:2008jh}.

Our manuscript is organized as follows.  We review our calculational approach
for determining the effects of the colored scalar on Higgs production
in Section~\ref{sec:calc}.  A more detailed discussion is presented in
our previous paper~\cite{Boughezal:2010ry}.  In Section~\ref{sec:nums}
we present the Tevatron bounds on these states, derived using the
model-independent constraints on Higgs production and decay via the
partonic process $gg \to h \to W^+W^-$.  We
conclude in Section~\ref{sec:conc}.

\section{Calculational details \label{sec:calc}}

We discuss here our calculational procedure for obtaining the
modifications in the Higgs production rate due to colored scalars.  We consider two example models of colored states that can modify the
SM prediction for gluon-fusion production of a Higgs boson: scalars in
the $(\bf{8},\bf{1})_{0}$ and $(\bf{3},\bf{1})_{0}$ representations.
The Lagrangians describing the interactions of these states with the
SM are given by
\begin{eqnarray}
\label{eq:Lfull}
{\cal L}^{adj} &=& {\cal L}_{SM}
         + \text{Tr}\left[D_{\mu} S D^{\mu} S\right]  
         - m_S^{'2} \,\text{Tr} \left[ S^2 \right]
         - g_s^2 \,G_{4S} \,\text{Tr} \left[ S^2 \right]^2
         -  \lambda_1 H^{\dagger}H \,\text{Tr} \left[ S^2 \right],
         \nonumber \\
{\cal L}^{fund} &=& {\cal L}_{SM}
         + \left(D_{\mu} S\right)^{\dagger} D^{\mu} S
         - m_S^{'2} \,S^{\dagger} S
         - \frac{1}{2}g_s^2 \,G_{4S} \,\left( S^{\dagger} S \right)^2
         -  \lambda_1 H^{\dagger}H \,S^{\dagger} S.
\end{eqnarray}
In the adjoint Lagrangian ${\cal L}^{adj}$, $S$ denotes the
matrix-valued scalar field $S = S^A T^A$, while in the fundamental
Lagrangian ${\cal L}^{fund}$, $S$ denotes a vector in color space with
three components.  $H$ indicates the Higgs doublet before electroweak
symmetry breaking, $v$ is the Higgs vacuum-expectation value, and
$D_{\mu}$ is the covariant derivative.  After electroweak symmetry
breaking, the Higgs doublet is expanded as $H = \left(0,(v+h)/\sqrt{2}
\right)$ in the unitary gauge.  The masses of the colored scalars become
$m_S^2 = m_S^{'2} + \lambda_1 v^2/2$.  The Feynman rules which
describe the scalar couplings to the Higgs boson $h$ and to gluons are 
easily obtained from Eq.~(\ref{eq:Lfull}).  The free parameters which
govern the scalar properties are $m_S$, $\lambda_1$, and $G_{4S}$.  We 
note that higher-order operators that break the $S \to -S$ symmetry 
present in the Lagrangians above, and which allow the scalar to decay,
can be obtained in
explicit models~\cite{Dobrescu:2007xf,Dobrescu:2007yp}.  We neglect
them here since we anticipate that they have little effect on the $gg
\to h$ production cross section.  We also note that a quartic-scalar
coupling is generated by QCD interactions even if it is set to zero at
tree-level.  At NNLO the quartic coupling must be included to obtain a
renormalizable result, as demonstrated in our previous work~\cite{Boughezal:2010ry}.  
We include this operator in the tree-level Lagrangian with a
coefficient scaled by $g_s^2$, the QCD coupling constant squared, to 
permit an easier power-counting of loops.

We calculate the production rate of the Higgs through NNLO in QCD
perturbation theory, which is necessary to reduce theoretical
uncertainties arising from variations of the renormalization and
facrorization scales in this process.  We perform this calculation
using an effective-theory strictly valid when the Higgs mass is less
than twice the scalar mass, and also less than twice the top mass, $m_h < 2m_{S,t}$.
In this limit, both the top quark and colored scalar can be integrated
out to obtain the following effective Lagrangian:
\begin{equation}
\label{eq:Leff}
{\cal L}^{eff} = {\cal L}_{QCD}^{n_l,eff}
         - C_1 \,\frac{H}{v} \,{\cal O}_1,
\end{equation}
where $C_1$ is a Wilson coefficient and the operator ${\cal O}_1$ is
\begin{equation}
{\cal O}_1 = \frac{1}{4} \,{G}^a_{\mu\nu} {G}^{a\mu\nu} \, .
\end{equation}
This form is valid for both the adjoint and fundamental scalar; only
$C_1$ changes.  This effective-theory has been extensively used to determine the
contribution of top-quark loops to the SM Higgs-gluon 
coupling, and to calculate the gluon-fusion cross section for Higgs
production at hadron
colliders~\cite{Shifman:1979eb,Inami:1982xt,Dawson:1990zj,Djouadi:1991tka,Kniehl:1995tn,Kramer:1996iq,Chetyrkin:1997un,Harlander:2002wh,Anastasiou:2002yz,Ravindran:2003um}. It
has also been used to determine the effect of a fourth generation of
fermions on the production rate at the Tevatron~\cite{Anastasiou:2010bt}.  
When normalized to the full $m_t$-dependent leading-order result, the 
effective-theory reproduces the exact NLO result obtained in
Ref.~\cite{Djouadi:1991tka,Spira:1995rr} to better than 1\% for $m_h <
2m_t$ and to 10\% or better for Higgs boson masses up to 1 TeV. For
scalars in the adjoint representation, this approximation has been
studied against the exact NLO calculation~\cite{Bonciani:2007ex}, 
and is again accurate to the $1-2$\% percent level for $m_h \leq
2 m_S$.  The deviation reaches a maximum of 10\% for Higgs masses much 
heavier than the scalar, except very near the threshold $m_h \approx 2
m_S$.  We adopt this normalized effective-theory as the basis for our
analysis.  A detailed derivation of the Wilson coefficient for the
adjoint scalar was presented in Ref.~\cite{Boughezal:2010ry}.  We
have derived $C_1$ for the fundamental scalar, and present it in Eq.~(\ref{eq:wilson}) of the Appendix.

Before presenting numerical results, we explain what contributions we include in the gluon-fusion cross
section and gluoinic decay width of the Higgs, both of which are
modified in the presence of the scalar.  The leading-order amplitude for the $gg \to h$ process takes the form
\begin{equation}
\label{eq:amplitude}
\mathcal{A}^{LO} = \mathcal{A}^{LO}_t+ \mathcal{A}^{LO}_b + \mathcal{A}^{LO}_S,
\end{equation}
where the subscripts $t,b,S$ respectively denote the top, bottom, and
scalar contributions.  The amplitude for $h \to gg$ has the identical
structure.  Upon squaring this amplitude, interferences
between the contributions of each particle are obtained.  We denote by
$\sigma_{t+S}$ the terms obtained by squaring together the top and
scalar amplitudes, and keeping both the interference term and the
pieces from each separate particle squared.  We let
$\sigma_{tb},\sigma_{Sb}$ denote the interferences between the
bottom-quark amplitude with the top and the scalar pieces.  
For the cross section at the $n$-th order in perturbation theory, we 
use the following expression:
\begin{equation}
\label{eq:framework}
\sigma^n = \sigma_{t+S}^{LO}(m_t,m_S) \,
K^n_{EFT}+\sigma_{Sb}^{LO}(m_S,m_b)+\sigma_{tb}^{LO}(m_t,m_b)+\sigma_{bb}^{LO}(m_b).
\end{equation}
$K^n_{EFT}$ denote the ratio of the $n$-th
order result for the cross section over the LO
result, with both quantities computed in the effective-theory defined in
Eq.~(\ref{eq:Leff}).  The expression multiplying the $K$-factors is
the LO cross section maintaining the exact dependence on the scalar and top-quark
masses. The remaining terms account for the scalar-bottom
interference, the top-bottom interference, and the bottom-squared
contribution at LO with their exact mass dependences. Various
electroweak corrections which modify the SM contribution at the
percent
level~\cite{Aglietti:2004nj,Actis:2008ug,Anastasiou:2008tj,Keung:2009bs,Brein:2010xj}
are not known for the scalar, and for consistency are neglected.  As
discussed above, this calculational framework yields results usually
accurate to the percent level, and to 10\% at worst.  For the SM it
gives results differing by only a couple of percent from the official
predictions utilized by the Tevatron
collaborations~\cite{Anastasiou:2008tj,deFlorian:2009hc}, and also
closely matches the prescription for LHC cross sections adopted by ATLAS and CMS~\cite{LHCHiggsCrossSectionWorkingGroup:2011ti}.  To calculate the 
partial width of the Higgs boson into gluons, we use HDECAY~\cite{Djouadi:1997yw} to
calculate its SM partial width.  We then scale its result by the ratio of the amplitude in Eq.~(\ref{eq:amplitude}) squared over that in the SM squared, together with a factor accounting for the different Wilson coefficients in the SM and in the presence of the scalar.  This partial width is then used with the other outputs of HDECAY to form the $h \to W^+W^-$ branching ratio.

\section{Numerical results \label{sec:nums}}

We now explore what regions of scalar parameter space are excluded by
the model-independent search of the CDF and D0 collaborations for the
process $gg \to h \to W^+W^-$.  We study the
adjoint and fundamental representations separately.  The addition of a
scalar to the spectrum induces two competing effects on the Tevatron
signal.  The scalar tends to increase the gluon-fusion production
cross section throughout the parameter space.  However, it also increases
the partial decay width of the Higgs into gluons, which decreases the
branching ratio into $W^+W^-$.  To
derive the allowed region of scalar parameter space we
calculate $\sigma_{gg \to h} \times BR(h \to WW)$, and demand that it
be less than the Tevatron limit on this quantity from
Ref.~\cite{Aaltonen:2010sv}.  In addition, to avoid strong couplings
and a breakdown of our perturbative analysis, we impose the additional
constraint $\Gamma_{total}/m_h < 1/5$ on the total width of the Higgs.  This constraint prevents the ratio $\lambda_1 v^2 / m_S^2$ appearing in the Wilson coefficient, where $v$ is the vacuum-expectation value of the Higgs, from becoming too large.

We use MSTW parton distribution functions~\cite{Martin:2009iq}
extracted through NNLO in QCD when
evaluating the cross section in Eq.~\ref{eq:framework}.  For the
bottom quark we use the pole mass $m_b = 4.75\, {\rm GeV}$; this leads
to slightly more conservative bounds than those obtained with the
$\overline{MS}$ mass.  The cross section is evaluated using the scale
choice $\mu_R=\mu_F =m_h/2$.  In the scalar sector, we must set the parameters $\lambda_1$,
$G_{4S}$, and $m_S$.  To determine the reasonable range of $G_{4S}$,
we perform a renormalization-group analysis and demand that this coupling does not encounter a Landau pole until a
cutoff $\Lambda = 10\, {\rm TeV}$.  This was discussed in detail in
Ref.~\cite{Boughezal:2010ry}, to which we refer the reader for further
details.  For the adjoint scalar, this leads to the approximate
condition $G_{4S}(v) < 1.5$, while for the fundamental we find $G_{4S}(v) <
2.5$.  The value of $G_{4S}$ does not significantly modify the bounds
we derive.  While we set $G_{4S}(v)=1$ in our study here, we have checked that
other values in the allowed region do not modify our bounds by more
than 5\%.  The cross section depends primarily on $\lambda_1$ through
the ratio $\lambda_1/m_S^2$, as discussed in
Ref.~\cite{Boughezal:2010ry} .  The constraints on $m_S$ depend
strongly on the value of $\lambda_1$ chosen.  However, there is no
symmetry reason to expect a small value for this  coupling, 
indicating a 'natural' value $\lambda_1 \approx 1$.  We note that
since the scalars considered here will not significantly alter
predictions for the precision electroweak observables measured at LEP
and SLC, the Higgs mass is also expected to be in the approximate
range $m_h < 200$ GeV.  Results for other values
$\lambda_1^{o}$, but keeping the mass fixed at $m_S$, can be
approximately obtained by studying the results at a mass $m_S^{o}$
given by $\lambda_1/m_S^2 = \lambda_1^{o} /(m_S^{o})^2$.  To avoid
three-dimensional plots we simply set $\lambda_1 = 1$ in our analysis,
and note that results for other values can be obtained by scaling
$\sigma_{gg \to h} \times BR(h \to WW)$ as indicated.  The only
remaining parameters are $m_h$ and $m_S$.  We present our results as
exclusion regions  in this two-dimensional space.

\begin{figure}[ht]
   \centering
   \includegraphics[width=0.55\textwidth,angle=90]{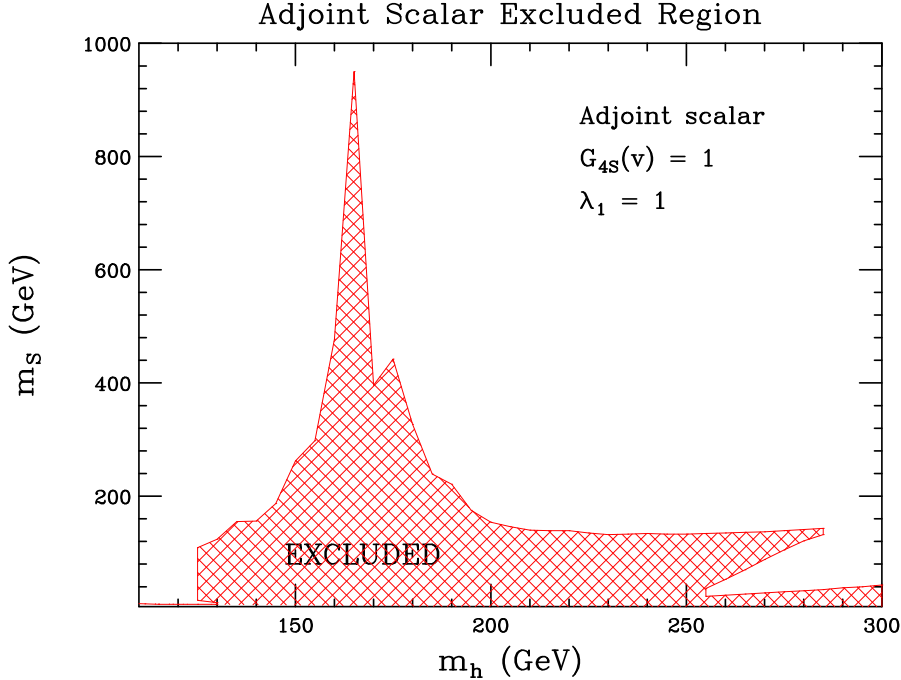}
   \vspace{-0.2cm}
   \caption{Bounds on the $(m_h,m_S)$ parameter space for an
     adjoint scalar.  The excluded values are denoted by red
     hatching.  The choices of $G_{4S}(v)$ and $\lambda_1$ are
     discussed in the text. }
   \label{fig:adjoint}
\end{figure}

\begin{figure}[ht]
   \centering
   \includegraphics[width=0.55\textwidth,angle=90]{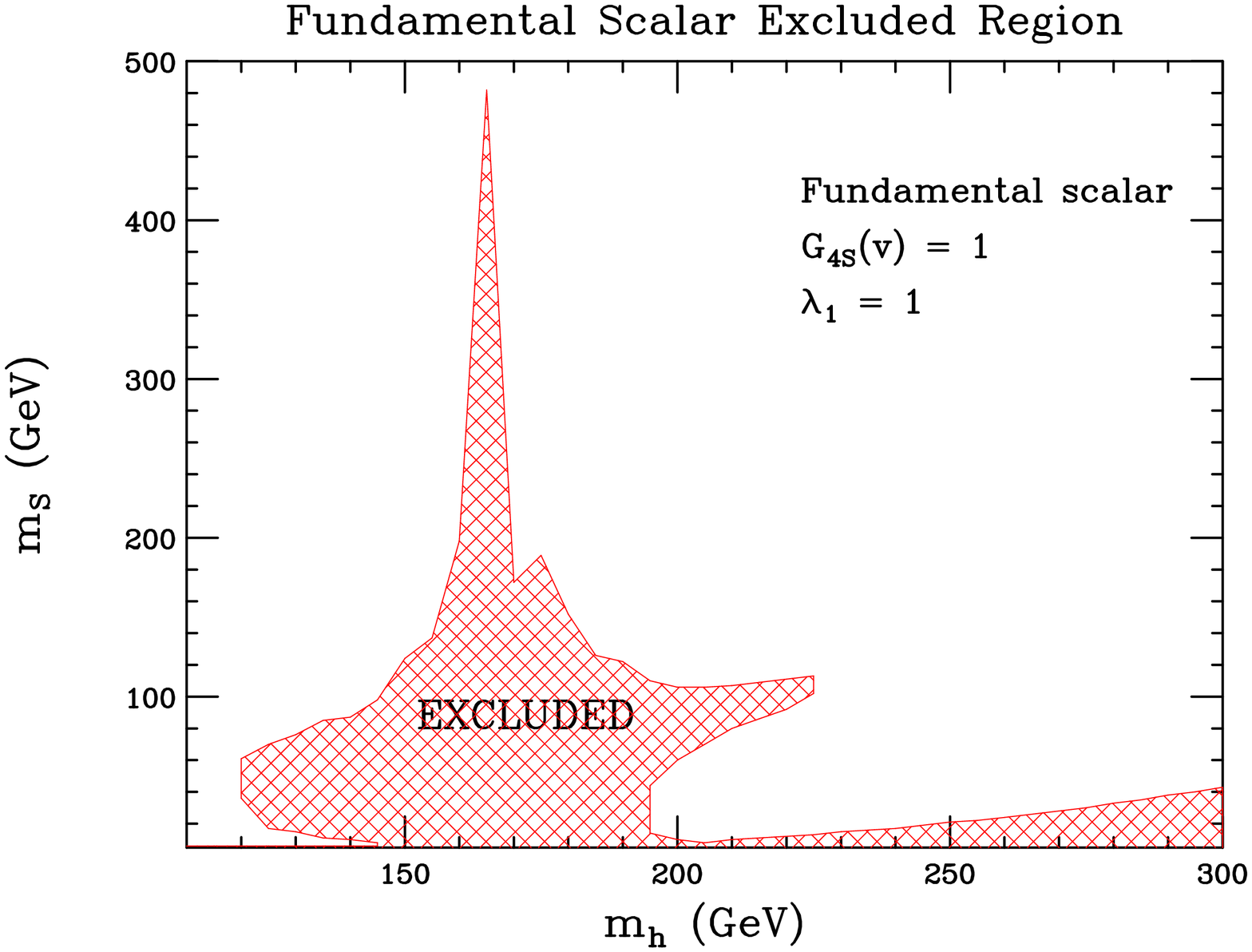}
   \vspace{-0.2cm}
   \caption{Bounds on the $(m_h,m_S)$ parameter space for a
     fundamental scalar.  The excluded values are denoted by red
     hatching.  The choices of $G_{4S}(v)$ and $\lambda_1$ are
     discussed in the text. }
   \label{fig:fundamental}
\end{figure}

The excluded regions for the adjoint and fundamental scalars are
presented in Figs.~\ref{fig:adjoint} and~\ref{fig:fundamental},
respectively.  We note that these are $95\%$ CL exclusion bounds. 
The strongest bounds occur when $m_h=165\,{\rm GeV}$ in
both cases. Since this search utilizes only the gluon-fusion
production mode, the SM Higgs is not excluded in this analysis, unlike
in the combined global analysis~\cite{:2010ar}.  Taking $m_S \to \infty$ gives the SM result for the production rate, and if the SM was excluded for a certain Higgs mass range in this analysis, even an extremely heavy scalar would not be allowed.  Adjoint-scalar masses
approaching 1 TeV are excluded for this Higgs mass, while
fundamental-scalar masses near 500 GeV are ruled out.  The constraints
are considerably weaker for other values of $m_h$, but adjoint-scalar masses
less than 130 GeV are ruled out for Higgs masses between 135 GeV and
250 GeV, while fundamental-scalar masses less than 100 GeV are
excluded for Higgs masses from 150 GeV to 190 GeV.  For comparison,
the direct-search limits on the adjoint scalar mass is estimated to be
280 GeV when the scalar decays primarily into $b\overline{b}$ at the
Tevatron~\cite{Dobrescu:2007yp}.  The direct search constraint is very
sensitive to the decay mode of the scalar.  However, it is also insensitive
to $m_h$ and $\lambda_1$; the two techniques for probing the scalar
parameter space are complementary.  The regions of light scalar mass are
allowed because $Br(h\to WW)$ drops quickly as $m_S$ is decreased.
However, scalar masses that are too light increase the gluonic partial width
to the point that $\Gamma_{total}/m_h$ becomes too large, leading to
the exclusion at small values of $m_S$. The
exclusion regions extending to large $m_S$ apparent in the figures regions at high $m_h$ arise
from the increase of the cross section near the threshold $m_h \approx
2 m_S$.  The allowed values of scalar masses found by our analysis are
presented in Tables~\ref{tab:adjoint} and~\ref{tab:fundamental}.  We note that
we consider only the Tevatron Higgs exclusion limit in determing these ranges.
Also, the very light allowed masses near $m_S \approx 10$ GeV should not be
taken too seriously due to the limitations of the effective-theory analysis.
While caution must be exercised in using our effective-theory framework for very
light scalar masses, we note that the deviation of
Eq.~(\ref{eq:framework}) form the exact next-to-leading order
calculation of Ref.~\cite{Bonciani:2007ex} was found to be less than
10\% for $m_h/m_S \leq 5$.  

We note that the bulk of the exclusion region for the adjoint scalar with a mass in 
the range $120-250$ GeV and the fundamental scalar with a mass in the range
$130-200$ GeV comes from the Tevatron bounds. The finite width constraint
has no effect in these regions. 
Also, in this dominant part of the excluded parameter space the effective-theory is working with high precision.
The properties of the width and the effective-theory limitations come in 
only in the tails extending to $m_h \geq 200$ GeV and very light scalar 
masses, where the Tevatron cross section becomes too small to constrain the
parameter space.

As a final comment, we also note that the validity of the
Tevatron exclusion limits have been actively debated within the past
year~\cite{debate}, due to the collaborations' treatment of theoretical
systematic errors.  We adopt the official results of the CDF and D0
collaborations in our analysis.

\begin{table}[ht]
  \begin{center}
    \begin{tabular}{|c|c||c|c|}
      \hline
      $m_h$ (GeV) & $m_S$ (GeV) & $m_h$ (GeV) & $m_S$ (GeV) \\
      \hline
      \hline
      110 & $10-\infty$ & 210 & $140-\infty$ \\ \hline
      115 & $9-\infty$ & 215 & $139-\infty$ \\ \hline
      120 & $9-\infty$ & 220 & $139-\infty$ \\ \hline
      125 & $9-16$, $109-\infty$ & 225 & $135-\infty$ \\ \hline
      130 & $9-12$, $124-\infty$ & 230 & $132-\infty$ \\ \hline
      135 & $155-\infty$ & 235 & $133-\infty$ \\ \hline
      140 & $156-\infty$ & 240 & $134-\infty$ \\ \hline
      145 & $187-\infty$ & 245 & $133-\infty$ \\ \hline
      150 & $262-\infty$ & 250 & $133-\infty$ \\ \hline
      155 & $298-\infty$ & 255 & $23-37$, $134-\infty$\\ \hline
      160 & $477-\infty$ & 260 & $25-53$, $135-\infty$ \\ \hline
      165 & $950-\infty$ & 265 & $27-70$, $136-\infty$ \\ \hline
      170 & $396-\infty$ & 270 & $29-88$, $137-\infty$ \\ \hline
      175 & $442-\infty$ & 275 & $31-105$,  $139-\infty$ \\ \hline
      180 & $327-\infty$ & 280 & $33-121$,  $141-\infty$ \\ \hline
      185 & $239-\infty$ & 285 & $35-132$,  $143-\infty$ \\ \hline
      190 & $221-\infty$ & 290 & $38-\infty$ \\ \hline
      195 & $175-\infty$ & 295 & $40-\infty$ \\ \hline
      200 & $154-\infty$ & 300 & $43-\infty$ \\ \hline
      205 & $146-\infty$ & $-$ & $-$ \\ \hline
        \end{tabular}
  \end{center}
  \caption{Allowed values of the adjoint scalar mass $m_S$ for each $m_h$,
    given the Tevatron bounds~\cite{Aaltonen:2010sv} and the
    constraint $\Gamma_{total}/m_h < 1/5$.  The choices of the
    parameters $\lambda_1$ and $G_{4S}$ are described in the text.  \label{tab:adjoint} }
\end{table}

\begin{table}[ht]
  \begin{center}
    \begin{tabular}{|c|c||c|c|}
      \hline
      $m_h$ (GeV) & $m_S$ (GeV) & $m_h$ (GeV) & $m_S$ (GeV) \\
      \hline
      \hline
      110 & $6-\infty$ & 210 & $10-80$, $107-\infty$ \\ \hline
      115 & $6-\infty$ & 215 & $11-86$, $109-\infty$ \\ \hline
      120 & $6-36$, $61-\infty$ & 220 & $12-92$, $111-\infty$ \\ \hline
      125 & $6-17$, $70-\infty$ & 225 & $13-102$, $113-\infty$ \\ \hline
      130 & $6-15$, $76-\infty$ & 230 & $15-\infty$ \\ \hline
      135 & $6-11$, $85-\infty$ & 235 & $16-\infty$ \\ \hline
      140 & $6-10$, $87-\infty$ & 240 & $17-\infty$ \\ \hline
      145 & $6-8$, $98-\infty$ & 245 & $19-\infty$ \\ \hline
      150 & $124-\infty$ & 250 & $21-\infty$ \\ \hline
      155 & $137-\infty$ & 255 & $22-\infty$\\ \hline
      160 & $198-\infty$ & 260 & $24-\infty$ \\ \hline
      165 & $482-\infty$ & 265 & $26-\infty$ \\ \hline
      170 & $172-\infty$ & 270 & $28-\infty$ \\ \hline
      175 & $189-\infty$ & 275 & $30-\infty$ \\ \hline
      180 & $152-\infty$ & 280 & $33-\infty$ \\ \hline
      185 & $126-\infty$ & 285 & $35-\infty$ \\ \hline
      190 & $122-\infty$ & 290 & $38-\infty$ \\ \hline
      195 & $14-44$, $110-\infty$ & 295 & $40-\infty$ \\ \hline
      200 & $10-60$, $106-\infty$ & 300 & $43-\infty$ \\ \hline
      205 & $9-70$, $106-\infty$ & $-$ & $-$ \\ \hline
        \end{tabular}
  \end{center}
  \caption{Allowed values of the fundamental scalar mass $m_S$ for each $m_h$,
    given the Tevatron bounds~\cite{Aaltonen:2010sv} and the
    constraint $\Gamma_{total}/m_h < 1/5$.  The choices of the
    parameters $\lambda_1$ and $G_{4S}$ are described in the text.  \label{tab:fundamental} }
\end{table}

\section{Conclusions \label{sec:conc}}

In this manuscript we have investigated the implications that Tevatron's
null search for the process $gg \to h \to W^+W^-$ has for extensions of the Standard Model.  We have considered two example
particles that substantially alter the gluon-fusion production of the
Higgs and its branching fraction into $W^+W^-$, and that also exist in
extensions of the SM: heavy scalars in the adjoint and fundamental
representations of the color gauge group.  We have outlined an
effective-theory computation of how the colored scalars modify Higgs
properties, and have also presented the Wilson coefficient for the
fundamental scalar needed for studies of this state through NNLO in QCD perturbation
theory.  Using the model-independent bounds on $\sigma_{gg \to h}
\times BR(h \to WW)$ presented in Ref.~\cite{Aaltonen:2010sv}, we have
  derived the excluded regions of scalar parameter space for both
  example states.  The constraints can be quite severe; for Higgs
  masses near $m_h \approx 165$ GeV, adjoint scalar masses approaching
  1 TeV are ruled out, while fundamental scalar masses near 500 GeV
  are not allowed.  These should be compared to the estimated direct
  search reach of $m_S \approx 280$ GeV for the adjoint
  scalar~\cite{Dobrescu:2007yp}.  Of course, the indirect bounds here
  depend significantly on $m_h$ and $\lambda_1$, which the
  direct-search limit does not.  However, it is also independent of
  the scalar decay, which plays a crucial role in the direct search.
  Throughout the natural range of $m_h$ and $\lambda_1$, the Tevatron results
  restrict the colored-scalar mass to be greater than $100-200$ GeV,
with the stronger restriction occurring for the adjoint representation.

The Tevatron null search for the Higgs boson has implications
beyond limiting the Standard Model parameter space.  It imposes severe
constraints on new theories that contain heavy
colored states which modify the Higgs-gluon interaction.  We have
demonstrated this by showing the strong limits obtainable on color adjoint
and fundamental scalars.  It would be interesting to investigate other
consequences of the Tevatron result for physics beyond the Standard Model.

\bigskip
\bigskip
\noindent
{\bf{\Large Acknowledgements}}
\bigskip

\noindent
Work supported by the US Department of Energy, Division of
High Energy Physics, under Contract DE-AC02-06CH11357.

\bigskip
\bigskip
\noindent
{\bf{\Large Appendix}}
\bigskip

\noindent
We present here the Wilson coefficient $C_1$ for the fundamental
scalar that appears in the effective Lagrangian of
Eq.~(\ref{eq:Leff}):
\begin{align}
 C_1 &= \; a\;\Bigl[-\frac{\lambda_1 \,v^2}{48 \,m_S^2}-\frac{1}{3}\Bigr]
         \;+ \;a^2\; \Bigl[\frac{\lambda_1\, v^2}{16\,m_S^2}
           \left(-\frac{3}{2}+\frac{G_{4S}}{3}\right) -\frac{11}{12}\Bigr]      
\nonumber \\
&+ \;a^3\;
\Bigl[-\frac{\left(225 x^6-72 x^4+131 x^2-228\right) \ln ^2(x)}{6144 (x-1)
x^2 (x+1)}
+\frac{\left(861 x^4+7435 x^2+684\right) \ln (x)}{9216\,x^2}
\nonumber \\
&+\frac{4608 \,L_S\, x^4-105 x^4+21888 \,L_S\, x^2-89129
x^2-2052}{27648 \, x^2}
+\,n_l\, \left(\frac{1}{288}(64 \,L_S\,+67)
+\frac{2 \ln (x)}{9}\right)
\nonumber \\
&
-\frac{\lambda_1\,v^2}{2\,m_S^2} \Bigl\{\frac{7}{96}(2 \,L_S\,-1) \,G_{4S}^2
+\left(\frac{424 \,L_S\,-823}{2304}+\frac{\ln(x)}{72}\right) \,G_{4S}
+\frac{(-6 \,L_S\,-71) \,n_l\,}{1728}
\nonumber \\
&
-\frac{\left(225 x^6-72 x^4+131 x^2-228\right) \ln^2(x)}{12288 (x-1) x^2
  (x+1)}
+\frac{\left(287 x^4-103x^2+228\right) \ln (x)}{6144\,x^2}
\nonumber \\
&
+\frac{4608 \,L_S\, x^4-2409 x^4-25152 \,L_S\, x^2+48247 x^2-2052}{55296\,x^2}
\Bigr\}
+\left(75x^6+26 x^4+7 x^2+76\right) \;\times
\nonumber \\
&
\left(\left(\ln (1+x)-\ln(1-x)\right) \frac{\ln^2(x)}{4096 \,x^3}
+\left(\text{Li}_2(-x)-\text{Li}_2(x)\right) \frac{\ln(x)}{2048 \,x^3}
+\frac{\text{Li}_3(x) - \text{Li}_3(-x)}{2048\,x^3}
\right.\nonumber \\
&\left.
-\frac{\lambda_1\,v^2}{2\,m_S^2} \Bigl\{\left(\ln (1+x)-\ln(1-x)\right) 
\frac{\ln^2(x)}{8192 \,x^3}
+\left(\text{Li}_2(-x)-\text{Li}_2(x)\right) \frac{\ln(x)}{4096\,x^3}
+\frac{\text{Li}_3(x) - \text{Li}_3(-x)}{4096 \, x^3} \Bigr\}
\right)\Bigr]\,.
\label{eq:wilson}
\end{align}
In these expressions, we have set $L_{i} = \mbox{ln}\left(m_{i}/\mu\right)$ for $i = (S,t)$,
$ x = m_t/m_S$ and $a=g_s^2/(4\pi^2)$.  As discussed in our previous
work~\cite{Boughezal:2010ry}, the combination $\lambda_1 v^2 /m_S^2$
is scale-invariant if only QCD-induced $\alpha_S$ corrections are
considered.  The Wilson coefficient therefore takes the same form in
both the $\overline{MS}$ and pole schemes through
next-to-next-to-leading order in perturbation theory.   In presenting our numerical results we interpret the masses of the top and scalar as pole masses.

\end{document}